# Radiative electron capture as a tunable source of highly linearly polarized x rays

M. Vockert,[1,2,*] G. Weber,[2,3] H. Bräuning,[3] A. Surzhykov,[4,5] C. Brandau,[3,6] S. Fritzsche,[2,7] S. Geyer,[8] S. Hagmann,[3] S. Hess,[3] C. Kozhuharov,[3] R. Märtin,[2] N. Petridis,[3] R. Hess,[3] S. Trotsenko,[3] Yu. A. Litvinov,[3] J. Glorius,[3] A. Gumberidze,[3] M. Steck,[3] S. Litvinov,[3] T. Gaßner,[3] P.-M. Hillenbrand,[3] M. Lestinsky,[3] F. Nolden,[3] M. S. Sanjari,[3] U. Popp,[3] C. Trageser,[3,6] D. F. A. Winters,[3] U. Spillmann,[3] T. Krings,[9] and Th. Stöhlker[1,2,3]

[1]*Institut für Optik und Quantenelektronik, Friedrich-Schiller-Universität Jena, Jena, Germany*
[2]*Helmholtz-Institut Jena, Jena, Germany*
[3]*GSI Helmholtzzentrum für Schwerionenforschung, Darmstadt, Germany*
[4]*Physikalisch-Technische Bundesanstalt, Braunschweig, Germany*
[5]*Fakultät für Elektrotechnik, Informationstechnik, Physik, Technische Universität Braunschweig, Braunschweig, Germany*
[6]*I. Physikalisches Institut, Justus-Liebig-Universität Gießen, Gießen, Germany*
[7]*Theoretisch-Physikalisches Institut, Friedrich-Schiller-Universität Jena, Jena, Germany*
[8]*Institut für Angewandte Physik, Goethe-Universität Frankfurt am Main, Frankfurt am Main, Germany*
[9]*Forschungszentrum Jülich, Jülich, Germany*

The radiative electron capture (REC) into the $K$ shell of bare Xe ions colliding with a hydrogen gas target has been investigated. In this study, the degree of linear polarization of the K-REC radiation was measured and compared with rigorous relativistic calculations as well as with the previous results recorded for $U^{92+}$. Owing to the improved detector technology, a significant gain in precision of the present polarization measurement is achieved compared to the previously published results. The obtained data confirms that for medium-Z ions such as Xe, the REC process is a source of highly polarized x rays which can easily be tuned with respect to the degree of linear polarization and the photon energy. We argue, in particular, that for relatively low energies the photons emitted under large angles are almost fully linear polarized.

DOI: 10.1103/PhysRevA.99.052702

## I. INTRODUCTION

Radiative electron capture (REC) is known as the dominant electron-capture process in fast collisions of heavy, highly charged ions with light target atoms and has attracted a lot of attention since its discovery in the early 1970s [1–16]. In this process, the capture of an electron from a target atom into a bound state of a projectile is accompanied by the emission of a photon that carries away the excess energy and momentum. Since in light targets the electrons are only loosely bound compared to the kinetic energy of the projectile, their initial states can be treated as quasifree. This makes the REC process similar to the radiative recombination (RR), which can be considered as the time inverse of photoionization in ion-atom collisions [13–16]. Over the last decades RR and REC total and angular differential cross sections were studied in great detail both by experiment and theory, including high-Z systems and relativistic collision energies (for a comprehensive overview see [16]). Based on all these studies, REC in collisions with open $K$- and $L$-shell ions and light target atoms appears to be well described by theory with respect to total as well as angular differential cross sections. Consequently, this process has meanwhile been established for normalization purposes of experimental data in the realm of atomic and nuclear physics [17–19]. A further important aspect of REC are the linear polarization properties of REC radiation, which has already been a subject of detailed theoretical investigations [20–22]. It was predicted that by varying the nuclear charge and collision energy of a projectile ion as well as the observation angle, a wide range of photon energies and degrees of linear polarization up to nearly 100% can be achieved. However, experimental studies of the REC polarization characteristics were hampered by the lack of efficient and precise polarimeter detectors for the hard x-ray regime (above 40 keV). As a consequence, to the best of our knowledge only one experiment has been reported so far [23]. In this pioneering experiment, performed for bare uranium ions in collision with a $N_2$ molecular target, the relevance of higher-multipole transitions has been demonstrated. Depending on observation angle and beam energy, these relativistic effects tend to lead to a depolarization of the radiation emitted. Nevertheless, a substantial degree of linear polarization of up to 80% has been found.

During the last decade the development of Compton polarimeters within the Stored Particle Atomic Research Collaboration (SPARC) [24] was motivated by the fact that many atomic physics processes lead to the emission of polarized x rays and are in turn also strongly dependent on the

---

polarization of the incoming particles [25,26]. Therefore, precise x-ray polarization measurements may offer new possibilities both for rigorous tests of atomic theory and also for the preparation and monitoring of polarized particle and photon beams, as was recently demonstrated for polarized electron beams [27–29] and for synchrotron radiation [30]. Moreover, as far as REC radiation is concerned, x-ray linear polarimetry was proposed as a tool for diagnosis of spin-polarized heavy-ion beams [31]. Such polarization diagnostics are one of the prerequisites for future experiments addressing atomic parity nonconservation [32], as may be addressed, e.g., at the upcoming Facility for Antiproton and Ion Research (FAIR). Fortunately, the introduction of improved, highly segmented semiconductor detector systems has significantly extended the applicability of Compton polarimetry so that it can now be regarded as a reliable tool in high-energy atomic physics [33]. Consequently, it is valuable for the study of subtle spin-dependent and relativistic effects in the realm of relativistic collisions involving ions at high Z [12,34,35].

In the present work we employ dedicated Compton polarimeters [36,37] developed by the SPARC Collaboration to provide REC (linear) polarization data with significantly increased precision compared to the previous study [23]. This enables a test of state-of-the-art, fully relativistic theory complementary to studies of the photon angular distribution. Moreover, we discuss for the case of mid-Z ions the REC process as a radiation source whose photon energy and degree of linear polarization is easily tunable over a broad range.

## II. THE EXPERIMENT

The data presented in this work was obtained at the internal gas target of the Experimental Storage Ring (ESR) at GSI, Darmstadt. Two measurement runs were performed using ion beams of bare xenon ($Xe^{54+}$) with kinetic energies of 30.9 and 150.3 MeV/u, respectively. The typical number of ions per injection was $10^8$, and the electron cooler of the ESR was applied to obtain a good beam quality, namely, a momentum spread of $\Delta p/p \leqslant 10^{-4}$ and a beam diameter of 2 mm [38]. Note, in case of the experiment at 30.9 MeV/u, the ions were injected at an energy of close to 100 MeV/u and decelerated to the final energy. A few seconds after injection, when stable beam conditions were reached, the internal gas target of the ESR was switched on. The stored ions were then passing through a hydrogen gas jet of a typical density of $10^{13}$ molecules per cm$^3$ [39] and a diameter of approximately 6 mm [40]. During each passage the ions may capture electrons from the target molecules, leading to a change in the projectile charge state and also the emission of x rays. While the down-charged ions were recorded with an efficiency close to 100% by a multiwire proportional chamber (MWPC) particle detector [41] located downstream of the target after the next dipole magnet, the x-ray emission originating from the interaction zone was recorded under an observation angle of 90° by two-dimensional (2D) position-sensitive Si(Li) detectors that were developed as dedicated Compton polarimeters.

By measuring the down-charged ions and x rays in coincidence we were able to select only those x rays that occur together with the capture of a target electron. Thus, the resulting spectra are dominated by features that can be attributed either to the REC process into the *K*, *L* and higher shells or to characteristic transitions, in particular, the Lyman lines (Ly), following the capture into excited states of the projectile, see Fig. 1(b).

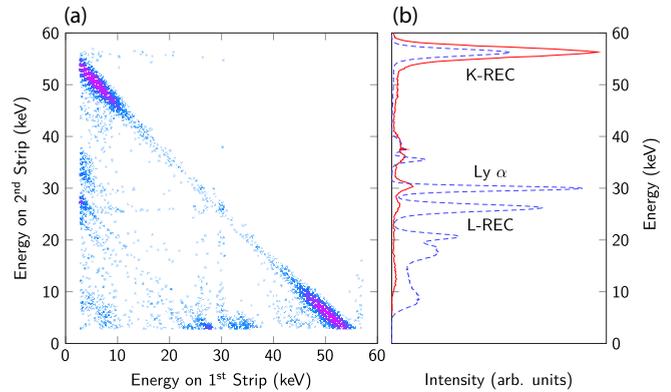

FIG. 1. X-ray spectrum recorded for the collision of $Xe^{54+}$ ions, moving with a kinetic energy of 30.9 MeV/u, with a H$_2$ target. (a) Scatter plot of energy depositions for those events, where exactly two coincident energy depositions were recorded by the front segments of the detector crystal. Diagonal lines in this plot designate Compton events of incident photons with a particular energy. (b) Spectrum of the incident photons absorbed in a single pseudopixel of the Compton polarimeter (dashed blue line) in comparison to the sum energy of the aforementioned double energy depositions (solid red line, scale increased for visibility). The most prominent feature of the latter spectrum is the K-REC peak at 56 keV. The peaks at lower photon energies are much less pronounced, as the efficiency of the detector as a Compton polarimeter drops off below roughly 45 keV.

During the first measurement at a kinetic beam energy of 150.3 MeV/u a prototype Si(Li) polarimeter of the SPARC Collaboration [36] was employed. A more recent polarimeter, which offers an improved energy resolution due to the lower electronic noise level of a cryogenic first stage of the preamplifiers [37], was used in the second beam time where the ion-beam energy was set to 30.9 MeV/u. Each detector consists of a single planar lithium-doped silicon crystal which is segmented into 32 horizontal strips on the front and 32 vertical strips on the back side. When combined, these strips form a structure of 1024 quadratic pseudopixels. The first detector is equipped with a 7-mm-thick crystal with an active area of 64 × 64 mm$^2$ (resulting in a strip width of 2 mm), while the crystal of the second detector has a thickness of 9 mm with an active area of 32 × 32 mm$^2$ (1 mm strip width). Each segment of the detector crystal is connected to its own charge-sensitive preamplifier and a subsequent readout chain, thus acting as an individual detector. More specifically, each strip provides information on the time ($\Delta t \approx 50$ ns) and energy (FWHM at 60 keV is 2–2.5 keV for the first detector and 0.8–1.2 keV for the second detector, with a cryogenic first stage of the preamplifiers) of each event.

As long as the number of interactions within one readout cycle is small, it is usually possible to unambiguously assign a pixel position to each energy deposition event within the detector crystal by combining the energy information from the front side and the back side. This provides 2D position resolution, which is the key to Compton polarimetry. More specifically, a Compton event results in energy depositions at the point of scattering (by the recoil electron) and in addition, at the position within the detector crystal where the scattered photon is absorbed. Combining energy, time, and position resolution then allows the reconstruction of the incident photon energy as well as the polar and azimuthal scattering angles. This is the basis for Compton polarimetry, which relies on the sensitivity of the scattering process to the degree and orientation of the incident photon linear polarization (see [42] for details).

Figure 1(a) displays a scatter plot of energy depositions for those events, where exactly two energy depositions were recorded by the front segments of the detector crystal. The data was obtained at the 30.9-MeV/u beam time using the Compton polarimeter with an improved energy resolution. The large number of events along the diagonal line at a sum energy of 56 keV corresponds to Compton scattering of incident K-REC photons inside the detector crystal. The low-energy part up to roughly 15 keV can be attributed to the recoil electrons, while the high-energy part corresponds to the scattered photons. The low electronic noise level of the improved detector allowed us to set the low-energy detection threshold to 3 keV, whereas the prototype detector with room-temperature preamplifiers required a significantly higher threshold of about 8 keV. The latter would, for the case of the 56-keV radiation, suppress a large portion of the recoil electrons and thus render the reconstruction of the associated Compton events impossible. The sum energy spectrum of the reconstructed Compton events is compared to those events where only one segment on each side of the detector recorded an event (i.e., photoabsorption of the incident photon) in Fig. 1(b). The latter spectrum is representative of the spectral distribution of the incident radiation, while the sum energy of Compton events has only the K-REC peak as a prominent feature. The peaks at lower photon energies are much less pronounced, as the efficiency of the detector as a Compton polarimeter drops off below roughly 45 keV. In the following we restrict ourselves to the (linear) polarization measurement of the most energetic type of REC radiation, namely, the K-REC.

## III. DATA ANALYSIS AND RESULTS

The degree of linear polarization was obtained by utilizing the polarization sensitivity of the Compton scattering process as described by the Klein-Nishina formula:

$$\frac{d\sigma}{d\Omega} = \frac{1}{2}r_0^2 \epsilon^2 \left( \epsilon + \frac{1}{\epsilon} - 2\sin^2\theta \cos^2\varphi \right). \quad (1)$$

Assuming scattering of an incident x ray off an electron at rest, the angular differential cross section depends on the ratio of the incident $E$ and scattered photon energies $E'$, $\epsilon = \frac{E'}{E}$, as well as the polar scattering angle $\theta$, and the azimuthal scattering angle $\varphi$.

The maximum anisotropy and, thus, polarization sensitivity with regard to the azimuthal scattering distribution occurs at polar scattering angles close to $90°$. Here, the azimuthal scattering distribution is roughly described by a $\cos^2\varphi$ dependence with respect to the electric field vector of the incident photon. Equation (1) can be adjusted to a beam of partially polarized photons by replacing $\cos^2\varphi$ with $\frac{1}{2}(1 - P) + P\cos^2(\varphi + \Delta\varphi)$, where $P$ corresponds to the degree of linear polarization and $\Delta\varphi$ accounts for a possible mismatch between the detector axis and the direction of the incident photon polarization.

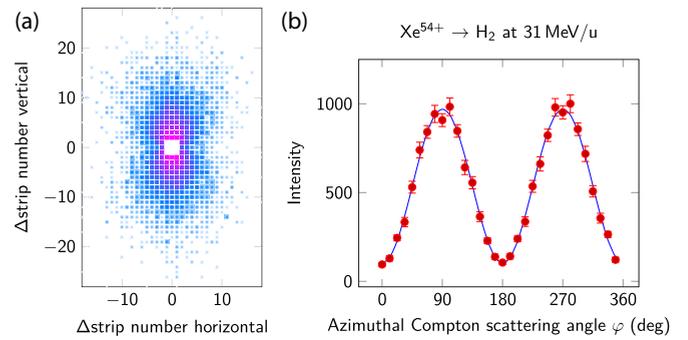

FIG. 2. (a) Position distribution of the Compton scattered K-REC photons with respect to the scattering position (0,0) in the polarimeter. Due to the high degree of linear polarization of the incident radiation, a clear anisotropy of the emission pattern is visible. (b) Azimuthal intensity distribution of the Compton scattered photons. The solid line is obtained by a least-squares adjustment of the Klein-Nishina formula and has been adapted to the partially polarized incident radiation of the experimental data.

As follows from the discussion above, both the degree of the linear polarization and the orientation of the polarization plane of the incident photons can be obtained from the scattered photon angular distribution with respect to the azimuthal angle $\varphi$; see [43,44] for a detailed discussion. While most conventional Compton polarimeter setups consist of a dedicated scatterer and one or more absorbers to detect the scattered photons, the 2D position-sensitive x-ray detectors used in the present study offer a much more efficient detection scheme, as each pixel can act as a scatterer and at the same time also as an absorber for the scattered radiation.

In our analysis, we have taken into account all Compton events associated with the K-REC peak having polar scattering angles of $\theta = (90 \pm 15)°$. The 2D position distribution of these Compton events inside the detector crystal is depicted in Fig. 2(a) for K-REC into $Xe^{54+}$ at 30.9 MeV/u. The strong anisotropy of the emission pattern indicates a high degree of linear polarization of the incident x rays. To obtain quantitative data, a least-squares adjustment of the modified Klein-Nishina equation with the degree of the incident radiation's linear polarization as a free parameter is applied to the azimuthal intensity distribution as shown in Fig. 2(b). The error bars of the experimental data points reflect the statistical uncertainties. However, the observed Compton scattering distribution is altered by several systematic effects, such as finite pixel size and limited energy resolution, which tend to lower the anisotropy as compared to Eq. (1). To correct for these systematic effects, the detector response was modeled by means of a Monte Carlo simulation based on the multiple-purpose photon transport code EGS5 [45]. It was previously demonstrated that such modeling is capable of reproducing all relevant detector characteristics [42,46]. In addition, we corrected the reconstructed linear polarization value for minor contributions due to background radiation and coincident hits of individual photons being falsely identified as Compton events in a similar way as was presented in [42]. The values of K-REC linear polarization resulting from the analysis are shown in Table I and depicted in Fig. 3, together with the theoretical predictions. These predictions have been obtained for the radiative recombination (RR) of a free electron into the bound ionic state. In this framework, both initial (continuum) and final (bound) electron states are described by the relativistic Dirac wave functions and a complete multipole expansion of the radiative field is taken into account [47,48]. Since the binding energies of the target electrons are negligible compared to the energy transfer during the capture into the projectile $K$ shell, we can assume that the RR process closely resembles the REC characteristics. In the nonrelativistic approximation, the photon emission of RR into the $K$ shell exhibits a $\sin^2 \phi$ emission pattern (with respect to the ion-beam axis) and 100% linear polarization independent of the observation angle $\theta$, i.e., the emission characteristics of a dipole. As shown in Fig. 3, this is a close approximation for the polarization of the K-REC into bare xenon at 30.9 MeV/u ($\beta = 0.25$), but already at a collision energy of 150.3 MeV/u ($\beta = 0.51$) a considerable depolarization due to relativistic effects is predicted by theory and also experimentally confirmed in this work. This is in contrast to studies of the REC photon angular distribution, where even at a relativistic collision energy of 197 MeV/u ($\beta = 0.56$) the emission pattern in the laboratory frame is effectively indistinguishable from the nonrelativistic $\sin^2$ shape [5]. Additionally, one can clearly see the strong shift between the angle in the emitter (Fig. 3, bottom) and laboratory frame (Fig. 3, top) due to the relativistic ion-beam energies. Also note that for high collision energies (>550 MeV/u for REC into the $K$ shell of $Xe^{54+}$) and at forward emission angles theory predicts a striking feature for REC radiation, namely, a negative linear polarization, i.e., a polarization perpendicular to the reaction plane, defined by the ion-beam axis and the momentum of the emitted photon [21]. In terms of the closely related process of photoionization, this corresponds to an ejection of the outgoing electron in the direction of the magnetic field vector of the incident photon.

TABLE I. Experimentally obtained values for the degree of linear polarization of K-REC radiation in comparison with theoretically predicted values for various collision systems and beam energies. Due to the improved polarimeter, the new values for bare xenon are much more precise than for bare uranium taken from [23].

| Collision system | Laboratory angle (deg) | Collision energy (MeV/u) | P experiment | P theory |
|---|---|---|---|---|
| $Xe^{54+} \to H_2$ | 90 | 30.9 | $0.999^{+0.001}_{-0.015}$ | 0.990 |
| $Xe^{54+} \to H_2$ | 90 | 150.3 | $0.956 \pm 0.017$ | 0.966 |
| $U^{92+} \to N_2$ | 90 | 400 | $0.79 \pm 0.08$ | 0.835 |
| $U^{92+} \to N_2$ | 60 | 400 | $0.61 \pm 0.12$ | 0.692 |
| $U^{92+} \to N_2$ | 60 | 190 | $0.72 \pm 0.05$ | 0.803 |
| $U^{92+} \to N_2$ | 60 | 132 | $0.83 \pm 0.05$ | 0.838 |
| $U^{92+} \to N_2$ | 60 | 98 | $0.85 \pm 0.07$ | 0.859 |

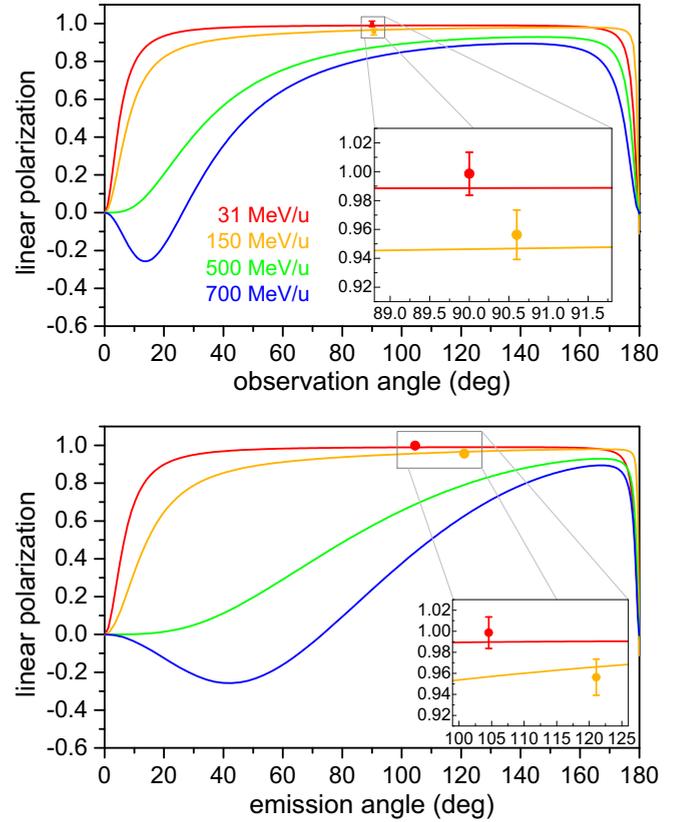

FIG. 3. Comparison of the experimentally observed linear polarization (filled dots) of the K-REC radiation with relativistic computations (solid lines) based on Dirac's equation as a function of the angle in the laboratory frame (top) and the emitter frame (bottom) for xenon ions at various collision energies (increasing values from upper to lower curve).

This feature has not been experimentally confirmed yet, as the kinetic energy of heavy-ion beams in the ESR is limited to roughly 400 MeV/u. With the upcoming High-Energy Storage Ring (HESR) of the future FAIR facility, the range of accessible energies for atomic physics experiments on stored ion beams will be significantly extended up to a few GeV/u [49]. In particular, for the case of $U^{92+}$ a beam energy of up to 5 GeV/u will be possible [24].

For REC into the $K$ shell of the heaviest ions a significant depolarization is found even at the lowest collision energies, as shown in Table I and Fig. 4 for the comparison of bare xenon and uranium projectiles. Therein results from rigorous relativistic treatment of the RR process [47,48], as well as experimentally obtained values, are given. As can be seen, the uncertainty of the determined degree of linear polarization was decreased by roughly a factor between 2 and 5 compared to the previous measurements due to the optimized detector setup.

Finally, it is also worth pointing out that the REC process can be utilized as a well-defined source of polarized x rays, tunable both in energy and the degree of linear polarization. Apart from the K-REC radiation, further REC lines are clearly resolved at distinct photon energies due to the capture into higher-lying shells, e.g., L-REC and M-REC. Moreover, the degree of linear polarization of the emitted light can be tuned over a broad range while covering an energy range from a few 10 keV up to several 100 keV by varying the observation angle, the ion species, and the collision energy, see Fig. 4. This might be of interest for measurements that do not require a high photon flux but a well-defined linear polarization, such as the calibration of advanced $\gamma$-ray polarimeters and tracking detectors (see, e.g., [50]). Here the REC process might provide a polarized photon source that has complementary features compared to synchrotron facilities, particularly for energies above 100 keV. Interestingly, the degree of linear polarization of the REC process is determined by basic experimental parameters such as beam energy, observation angle, and projectile nuclear charge.

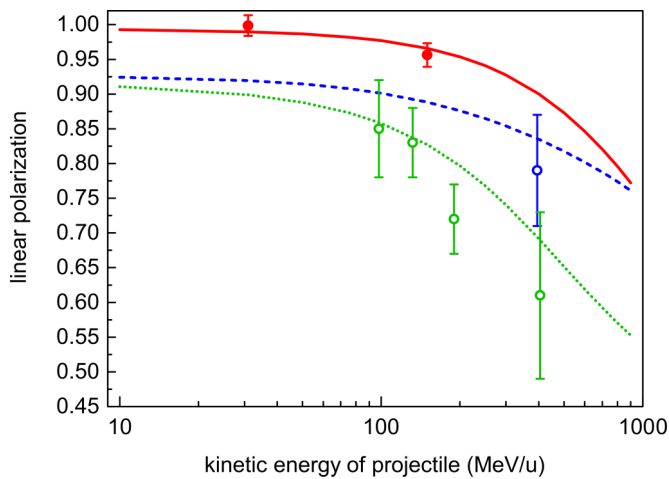

FIG. 4. Polarization as a function of the collision energy for xenon and uranium projectiles in the laboratory frame. Red line: theory values for Xe$^{54+}$ on H$_2$ under 90°; blue dashed line: theory values for U$^{92+}$ on N$_2$ under 90°; green dotted line: theory values for U$^{92+}$ on N$_2$ under 60°; filled red dots: experimentally obtained values presented in this work; open dots: experimentally obtained values taken from [23].

## IV. CONCLUSION

In summary, the linear polarization of the REC radiation has been measured for bare xenon in collisions with a molecular hydrogen target at beam energies of 31 and 150 MeV/u, respectively. Compared to the REC polarization measurements published before, an overall improvement of the relative precision by roughly a factor 2–5 has been achieved by applying improved Compton polarimeters. Experimental findings are in good agreement with state-of-the-art theory and prove that the REC process yields x-ray linear polarization up to nearly 100%. Both the photon energy as well as the degree of linear polarization can be tuned over a broad range by varying the energy and charge of the projectile ions and the observation angle.

## ACKNOWLEDGMENTS

We acknowledge financial support by the German Federal Ministry of Education and Research (BMBF) under Grants No. 05P15SJFAA and No. 05P15RGFAA, as well as from the European Research Council (ERC) under the European Union's Horizon 2020 research and innovation programme (Grant Agreement No. 682841 "ASTRUm").